\documentclass{article}
\usepackage{spconf,amsmath,graphicx,amsfonts,url,color,epstopdf}
\usepackage[linkcolor=black]{hyperref}

\title{STATISTICAL PARAMETRIC SPEECH SYNTHESIS USING GENERATIVE ADVERSARIAL NETWORKS UNDER A MULTI-TASK LEARNING FRAMEWORK}
\name{Shan Yang$^1$,
      Lei Xie$^1$,
      Xiao Chen$^2$,
      Xiaoyan Lou$^2$,
      Xuan Zhu$^2$,
      Dongyan Huang$^3$,
      Haizhou Li$^{3,4}$
}
\address{$^1$School of Computer Science, Northwestern Polytechnical University, Xi'an, China\\
         $^2$Language Computing Lab, Samsung R\&D Institute of China, Beijing, China\\
         $^3$Institute for Infocomm Research, A$^\star$STAR, Singapore\\
         $^4$Department of Electrical and Computer Engineering, National University of Singapore}

\begin{document}
\topmargin=0mm
%
\maketitle
\begin{abstract}
In this paper, we aim at improving the performance of synthesized speech in
statistical parametric speech synthesis (SPSS) based on a generative adversarial network (GAN).
In particular, we propose a novel architecture combining the traditional acoustic loss function and the
GAN's discriminative loss under a multi-task learning (MTL) framework. The mean squared error (MSE) is usually used to estimate the parameters of deep neural networks,
which only considers the numerical difference between the
raw audio and the synthesized one. To mitigate this problem, we introduce the GAN as a second task to
determine if the input is a natural speech with specific conditions. In this MTL framework, the MSE optimization improves the stability
of GAN, and at the same time GAN produces samples with a distribution closer to natural speech. Listening tests
show that the multi-task architecture can generate more natural speech that satisfies human perception than the conventional methods.
\end{abstract}
\section{INTRODUCTION}
\label{sec:intro}

Statistical parametric speech synthesis (SPSS) has attracted significant attentions since the successful use of hidden Markov models (HMMs)~\cite{yoshimura1999simultaneous,tokuda2000speech,zen2009statistical}.
In HMM based systems, Gaussian mixture model (GMM) was used to model
the hidden states of observations. Considering the limitations of the decision tree clustering procedure in modeling the complex context dependencies in HMM-based statistical parametric speech synthesis~\cite{ze2013statistical,ling2015deep}, deep neural networks (DNNs)
have been proposed for acoustic modeling, which can produce more natural synthesized speech~\cite{ze2013statistical,watts2016hmms}.
More recently, advanced estimation criteria and novel network architectures have been introduced to further improve the
performance of SPSS~\cite{ling2013modeling,kang2013multi,zen2014deep,fan2014tts,wangautoregressive}.

Since the purpose of training in the statistical methods is to maximize the likelihood or specifically to minimize the mean
square error (MSE) between the synthesized  (i.e., network outputs) and the original speech parameters in neural network based Text-to-Speech (TTS), the synthesized speech may achieve suboptimal human perceptual level. Hence there is an underlying reasonable-but-not-necessarily-optimal hypothesis that the most natural synthesized speech has the minimal value in the numerical loss, which may fall into the \emph{perceptual deficiency} problem.  In other words, the reduction in numerical errors may not necessarily lead to better perceived speech~\cite{wu2015deep}. In this paper, we propose to use generative adversarial networks (GANs)~\cite{goodfellow2014generative} to remedy this deficiency.

Significant efforts have been made to remedy the perceptual deficiency problem by improving the training criteria~\cite{yuchensequence, wu2015minimum, saito2017training,wu2016improving}. In~\cite{yuchensequence}, by incorporating the whole sequence parameters into training, the
sequence generation error (SGE) minimization was proposed to eliminate the mismatch between training and testing. Considering the independence of frames in DNNs, the minimum trajectory error training
was adopted to take into account the dynamic constraints from a wide acoustic context during training~\cite{wu2015minimum}.  In SPSS, the speech features must be invertible for reconstruction through a vocoder, and this rules out the use of many perceptual representations of speech that can not be reconstructed to speech waveform. Hence one solution to the \emph{perceptual suboptimality} issue is to bring more representative perceptual features into acoustic modeling~\cite{wu2015deep}. In~\cite{wu2015deep}, under a multi-task learning (MTL) framework, along with the invertible spectral feature used in the vocoder, extra perceptual representations of speech, e.g., spectro-temporal excitation pattern, were included as a second prediction target  in DNN-based SPSS. 

In this paper, we propose to use GANs to solve the perceptual deficiency problem in acoustic modeling. GAN is a powerful generative model that has been successfully
used in image generation~\cite{goodfellow2014generative,mirza2014conditional,Radford2015Unsupervised} and other tasks~\cite{pascual2017segan,yang2017midinet}. It consists
of a generator $G$, which is treated as an acoustic model in our framework to generate speech, and a discriminator $D$ for discriminating the generated speech and the genuine speech.
Specifically, the objective of $G$ is to capture the distribution of the natural speech, while $D$ aids the training of $G$ by examining the data generated by $G$ in reference to real data, and hence
helping $G$ learn the distribution that underpins the real data~\cite{goodfellow2014generative}. 
In our framework, GAN naturally addresses the perceptual deficiency problem: the updating of the generator is
not directly from the data samples, while it comes from the back propagation of the discriminator. This means $D$ can capture the essential difference between the natural speech and the synthesized speech and this `perceptual' difference is used to guide $G$, the generator. Considering the mode collapse problem of the generated samples in GAN~\cite{mirza2014conditional}, we take conditional linguistic features as a guidance to
control the generation process. Moreover, since the gradients of GANs are not stable, we also use the conventional MSE loss
function to stabilize the training process. More specifically, inspired by~\cite{saito2017training,wu2015deep,rosca2017variational,45882}, we combine the MSE loss with the GAN loss under an MTL framework. The objective experiments show that our framework has comparable performance in numerical loss compared to the baseline BLSTM-based TTS, while promisingly, the subjective listening experiments indicate that the proposed architecture achieves significant improvement. That is, the proposed GAN approach results in better perceptual speech quality.

\section{RELATED WORKS}
\label{sec:rel}

We notice that there are several recent attempts of using GANs to improve the quality of synthesized speech. In~\cite{kaneko2016generative},
GAN was treated as a post-filter for acoustic models to overcome the over-smoothing problem. Specifically, natural speech was used as a conditional guidance of GAN, which tries to reproduce the natural speech texture from the synthesized one.  In~\cite{hsu2017voice}, variational autoencoding Wasserstein GAN
(VAW-GAN) was proposed to build a voice conversion system from unaligned data, in which the GAN objective was incorporated into the decoder to improve the conditional variational autoencoder (C-VAE).

Our approach shares a similar idea with \cite{saito2017training}. In order to compensate the difference between the synthesized speech and the natural speech in acoustic modeling, an Anti-Spoofing Verification (ASV) module (like the discriminator in GAN) was introduced to distinguish between the natural and the synthetic speech.  The speech generator has no difference with a typical neural network acoustic model~\cite{ze2013statistical}, i.e., learning a non-linear mapping from linguistic features to speech parameters, but the ASV discrimination loss was combined with the minimal generation error (MGE) loss, under an MTL framework, to train the network.


It is noted that our approach is different from~\cite{saito2017training} in terms of motivation and  implementation. Instead of addressing the over-smoothed problem with additional ASV constraint as compensation, we propose to use GANs to directly produce speech samples with closer distribution to natural speech from a \emph{uniformly random noise distribution}. In other words, the input of our speech generator is random noise, while linguistic features are introduced to both the generator and the discriminator as conditions. As such, the prior uniformly random noise distribution creates new samples that approximate the training data distribution, and it brings diversity with conditions to the synthesized speech from the generator while the linguistic conditions add direct linguistic-discriminative information to the discriminator. On the other hand, as the Nash equilibrium is hard to achieve in network estimation, the training process becomes unstable during the adversarial game. To tackle this problem, we take other optimization methods, such as variational auto-encoder~\cite{rosca2017variational} or MSE, to restrain the process. Finally, in our implementation,  we use state-of-the-art BLSTM network as a benchmark, which can produce speech with much better quality than feed-forward networks used in \cite{saito2017training}.


\section{GAN-BASED MULTI-TASK LEARNING}
\label{sec:framework}

\begin{figure}[tb]
  \centering
  \centerline{\includegraphics[width=8.5cm]{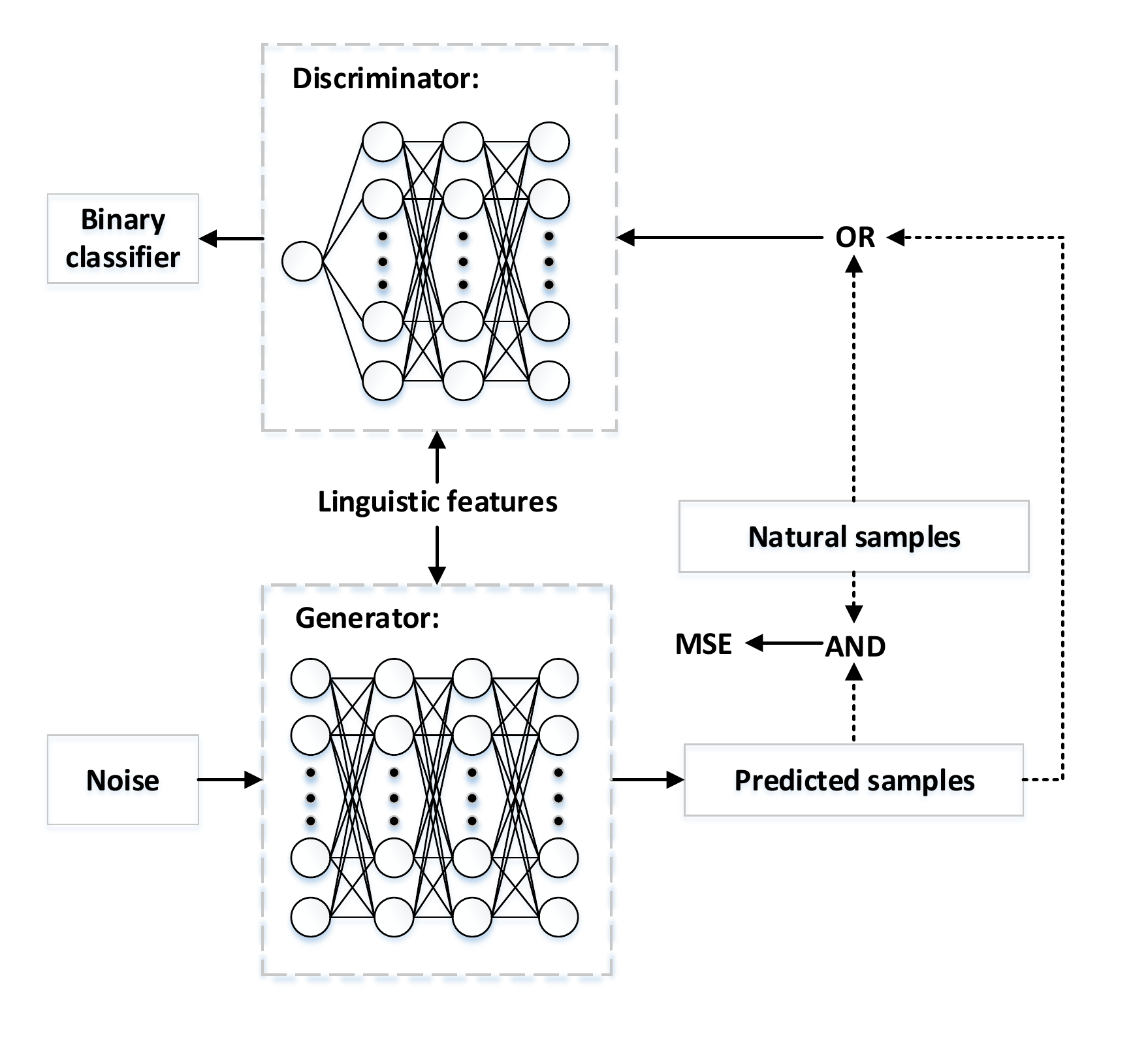}}
%
\caption{System diagram of GAN-based multi-task learning framework.}
\label{fig:system}
\end{figure}

Fig.~\ref{fig:system} illustrates the architecture of the proposed GAN-based MTL framework, which consists of a \emph{Generator}
and a \emph{Discriminator}. In the training process, different from \cite{saito2017training}, we use random noise as the input of \emph{Generator}, and introduce the linguistic
features to each hidden layer as the conditional information. Then the \emph{Generator} can produce the synthesized speech, with which the \emph{Discriminator}
can distinguish between the synthesized speech and the natural speech under the same conditions.
The estimation of this process is made up of two aspects: 1) For \emph{Discriminator}, the \emph{OR} operator means that the synthesized samples and
natural samples are alternately used to train a binary classifier (whether synthesized or genuine speech). 2) As for \emph{Generator}, the \emph{AND} operator is related to the MSE calculation
and its optimization, and meanwhile the \emph{OR} operator signifies that the discriminant error will also affect the estimation. In the synthesis stage, given a random
noise and specific linguistic features, we can easily generate speech from \emph{Generator} using the forward direction. We will describe the framework in details in the following.

\subsection{Generative Adversarial Networks}
\label{ssec:gan}

GAN is a generative model that can learn a complex relationship between random
noise input vector $z$ and output parameters $y$ by an adversarial process~\cite{goodfellow2014generative}.
The estimation of GANs consist of two models: a generative model $G$ that captures
the data distribution from random noise $z$, and a discriminative model $D$ that
maximizes the probability of correctly discriminating between the \emph{real} examples
and \emph{fake} samples generated from $G$.

In this adversarial process, the generator tends to learn a mapping function $G(z)$
to fit the real data distribution $p_{data}(x)$ from a uniformly random noise distribution $p_z(z)$, while the purpose of discriminator is to perfectly judge whether the sample is from
$G(z)$ or $p_{data}(x)$. So the $G$ and $D$ are both trained simultaneously in the
two-player min-max game with value function:
\begin{multline}
loss_{gan} = \min \limits_G \max \limits_D \mathbb{E}_{x{\sim}p_{data}(x)}[{\rm log}D(x)] \\
+ \mathbb{E}_{z{\sim}p_{z}(z)}[{\rm log}(1-D(G(z)))].
\label{eq1}
\end{multline}

In the above generative model, the modes of generated samples cannot be controlled
because of the weak guidance. So the conditional generative adversarial network (CGAN)~\cite{mirza2014conditional}
is proposed to direct the generation
 by considering additional information $y$. Then the loss function can be expressed
 as
 \begin{multline}
loss_{cgan} = \arg\min \limits_G \max \limits_D \mathbb{E}_{x{\sim}p_{data}(x)}[{\rm log}D(x|y)] \\
+ \mathbb{E}_{z{\sim}p_{z}(z)}[{\rm log}(1-D(G(z|y)|y))].
\label{eq2}
\end{multline}

\subsection{Multi-Task Learning with GANs in SPSS}
\label{ssec:multi-task}

In the traditional acoustic model for SPSS, we usually minimize the MSE between the predicted parameters $X_{model}$ and the natural speech
$X_{real}$ during the estimation. The objective can be written as
 \begin{equation}
loss_{mse}  = \arg\min \frac{\sum_{i=1}^n(X_{real,i} - X_{model,i})^2}{n}.
\label{eq3}
\end{equation}

As Eq.~(\ref{eq3}) shows, the numerical difference (in terms of MSE) is only concerned in the estimation,
and the numerical error reduction may not necessarily lead to perceptual improvement on the synthesized speech~\cite{wu2015deep}. To solve this problem, we propose to use GANs to learn the essential differences between the synthesized speech and the natural speech through a discriminative process.

 GAN is able to generate data rather than estimate the density function. Due to the model collapse problem in the generative model in GAN~\cite{mirza2014conditional}, we propose the following generator loss function  in order to guide GAN to converge to optimal solution such that the generative model produces desired data:
 \begin{multline}
 \mathbb{E}_{z{\sim}p_{z}(z)}[G(z|y))- X_{real}]^2 + \\
 \mathbb{E}_{z{\sim}p_{z}(z)}[{\rm log}(1-D(G(z|y)|y))],
\label{eq4G}
\end{multline}
where $X_{real}{\sim}p_{data}(x)$, and $X_{model}$ is generated by the generator G using uniformly random noise $z$ under condition $y$. Combining
Eq.~(\ref{eq2}) and Eq.~(\ref{eq4G}), the final objective of our MTL framework is:
 \begin{multline}
loss_{multi}  = \arg\min \limits_G \max \limits_D \mathbb{E}_{x{\sim}p_{data}(x)}[{\rm log}D(x|y)] \\
+ \mathbb{E}_{z{\sim}p_{z}(z)}[G(z|y))- X_{real}]^2 \\
+ \mathbb{E}_{z{\sim}p_{z}(z)}[{\rm log}(1-D(G(z|y)|y))].
\label{eq4}
\end{multline}

We treat the linguistic features as additional vector $y$, and make the input noise
$z$ obey a uniform distribution in the interval [-1,1]. Then our framework can generate the speech
$X_{model}$ by $G(z|y)$, and the $loss_{mse}$ and $loss_{cgan}$ are estimated at the same time
during training. Note that the input of our speech generator is uniformly random noise and linguistic features are used as conditions for both the generator and the discriminator, which in different from~\cite{saito2017training}.

Since the effective likelihood of GAN is unknown and intractable~\cite{rosca2017variational}, several auto-encoder GAN variants use zero-mean Laplace distribution $\exp(-\lambda||x-G(z)||_1)$~\cite{berthelot2017began,nguyen2016plug} to solve the problems. In order to directly show the likelihood of these variants, we can simply set $\lambda = 1$ and replace the $L1$ reconstruction
loss with $L2$ norm, and then we can get the MSE format as traditional methods. That is to say, we can take other explicit
likelihood (e.g., MSE) to solve the intractable inference of GANs. The $L1$ reconstruction loss will be investigated in the near future.

\subsection{Phoneme Discrimination for GANs}
\label{ssec:phoneme-discrimination}

\begin{figure}[tb]
 \centering
  \centerline{\includegraphics[width=8.5cm]{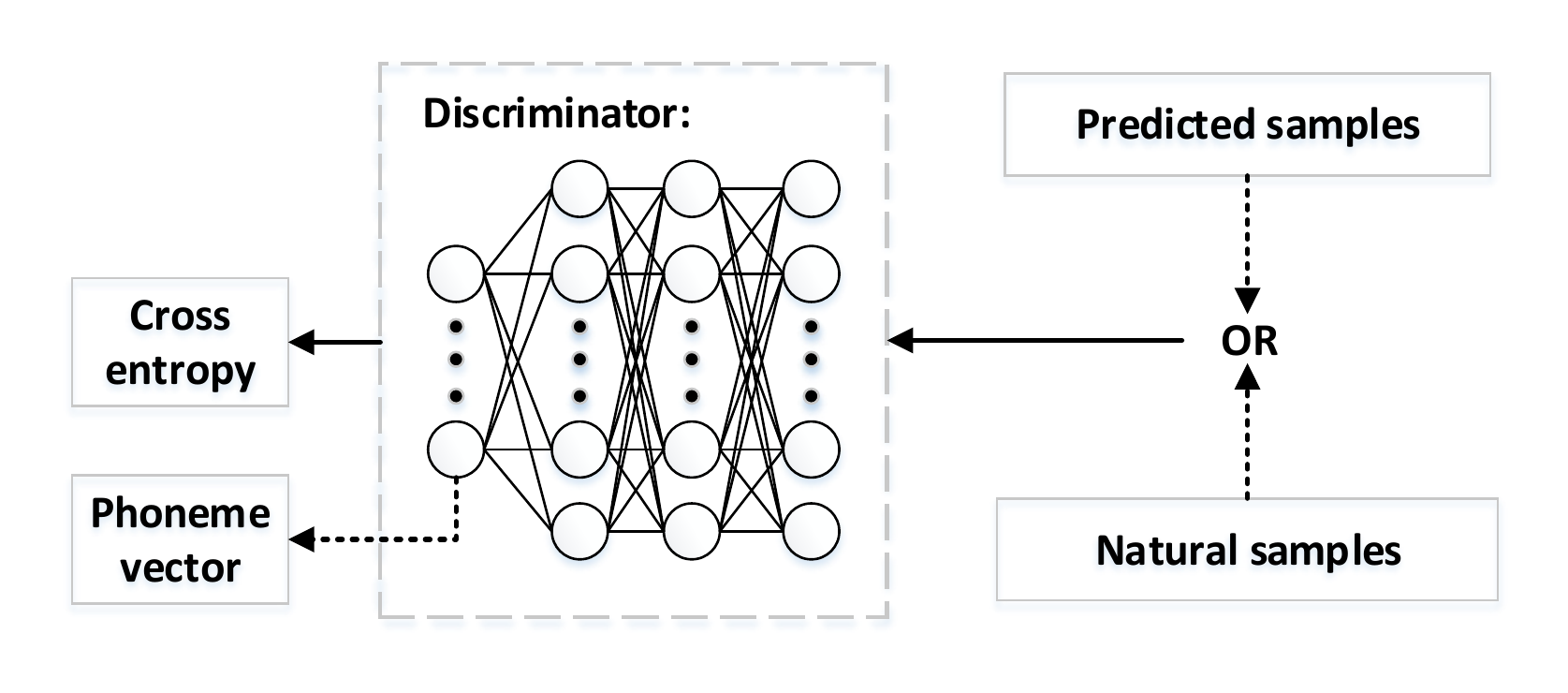}}
%
\caption{The discriminator with phoneme information.}
\label{fig:newD}
\end{figure}

In Section~\ref{ssec:multi-task}, the discriminator is a binary classifier to judge
whether the data $x$ is from $G(z)$ or $p_{data}(x)$ under the condition $y$. We also try to use phoneme information to guide the discrimination process in our multi-task framework, as shown in Fig.\ref{fig:newD}.

Assume $label$ is a one-hot encoded vector representing the phoneme class, which is
the category of both \emph{fake} and \emph{real} samples for $D$. Then our goal is to minimize the
cross entropy (CE) for the \emph{real} and to maximize this loss for the \emph{fake}, and the latter one means
that we do not know which phoneme the \emph{fake} belongs to. So the target function of GANs in
$loss_{multi}$ can be updated with
\begin{multline}
\arg\min \mathbb{E}_{x{\sim}p_{data}(x)}[D_{CE}(x|y,label)] \\
- \mathbb{E}_{z{\sim}p_{z}(z)}[D_{CE}(G(z|y)|y,label)].
\label{eq5}
\end{multline}

We obtain the new loss function considering the phoneme classification as follows.
 \begin{multline}
loss_{multi-pc}  = \arg\min \mathbb{E}_{x{\sim}p_{data}(x)}[{\rm log}D_{CE}(x|y,label)] \\
+ \mathbb{E}_{z{\sim}p_{z}(z)}[G(z|y))- X_{real}]^2 \\
- \mathbb{E}_{z{\sim}p_{z}(z)}[D_{CE}(G(z|y)|y,label)].
\label{eq6}
\end{multline}

\section{EXPERIMENTS}
\label{sec:experiments}

\subsection{Experimental Setup}
\label{ssec:setups}

In the experiments, a Chinese speech corpus was used to evaluate the performance of our approach. The corpus consists of about 10,000 utterances from
a single female speaker. We randomly selected around 8,000 sentences for network training,
1,000 utterances for model validation and another 1,000 for testing. Each speech waveform
was sampled at 16 kHz, and we used WORLD~\cite{morise2016world} (D4C edition~\cite{morise2016d4c})
to extract 60-dimensional Mel-Cepstral Coefficients (MCCs), 1-dimensional band aperiodicities (BAP)
and $F_0$ in log-scale in 5-ms step. So the final acoustic features were 63-dimensions including one
extra binary voiced/unvoiced flag. As for the text, we made a complex text analysis module to
get 138-dimensional linguistic features, including phoneme information, prosody boundary labeling, part
of speech tagging, state information and corresponding position index.

To benchmark the performance of the GAN-based MTL framework, we compared four systems, listed as follows.
\begin{itemize}

\item \textbf{BLSTM}: We used bidirectional long short-term memory (BLSTM) based acoustic model as the baseline, which contained three feed-forward
layers with 512 nodes/layer, followed by two BLSTM layers with 512 cells and a fully-connected output layer.
\item \textbf{GAN-based MTL (GAN)}: The proposed framework shown in Fig.~\ref{fig:system}. For the generator $G$, we also used three feed-forward and two BLSTM layers
corresponding to the baseline. But the input was replaced with 200-dimensional random noise under the
[-1,1] uniform distribution. The linguistic features were added to the output of each hidden
layer in $G$ as conditions. As for the discriminator, two convolutional layer were used with $5*5$ filter
shape, and $LReLU$ was treated as the activation function followed by batch
normalization~\cite{kaneko2016generative,Radford2015Unsupervised}. Besides, there was a fully-connected layer after the convolutional architecture and a binary classification layer in the end. The linguistic
conditions were also introduced to all hidden layers in $D$.
\item \textbf{ASV as a second task~\cite{saito2017training} (ASV)}: We realized the ASV approach with $\omega_D=1$. The network architecture was the same as \textbf{BLSTM}.
\item \textbf{GAN with phoneme classification (GAN-PC)}: The same \textbf{GAN} model architecture was used, except that the output layer of $D$ became a 63-category phoneme classification task, as described in Section~\ref{ssec:phoneme-discrimination}.
\end{itemize}

All the above methods were optimized using Adam optimizer~\cite{kingma2014adam,arjovsky2017wasserstein}, and we implemented all the systems with TensorFlow~\cite{abadi2016tensorflow}.

\subsection{Objective Evaluation}
\label{ssec:obeva}

We first conducted the objective measure to evaluate the performance of the four systems on the testing data.
Specifically, Mel-cepstral distortion (MCD) was used to evaluate the distortion of spectrum, and  RMSE was introduced to calculate the $F_0$ error. Besides, the V/UV error rate was also used to present
the accuracy of the voice/unvoice flag judgements.

\begin{table}[t]
        \footnotesize
        \centerline{
          \begin{tabular}{|c|c|c|c|}
            \hline
                Methods & {MCD} (dB) & {$F_0$ RMSE (Hz)} & {V/UV (\%)} \\
            \hline \hline
               BLSTM      & $4.624$ & $18.544$ & $6.447$ \\
               ASV~\cite{saito2017training}     & $4.670$ & $18.871$ & $6.562$ \\
               GAN     & $4.633$ & $18.678$ & $6.492$ \\
               GAN-PC         & $4.628$ & $18.616$ & $6.464$ \\
            \hline
          \end{tabular}
        }
        \caption{\label{tab:objective} {\it Objective evaluation results. }}
\end{table}

Table~\ref{tab:objective} shows the objective results. As shown, there seems to be no remarkable differences among
these systems. Since one purpose of our framework was to compensate the deficiency of traditional numerical
loss function by GANs, these numerical measures may not be suitable for evaluation because of the internal squared
error~\cite{kaneko2016generative}.  But in another aspect, we can find that the proposed MTL framework can compensate the instability and the mode collapse issues of GANs and generate stable and diverse speech with
the help of traditional loss function. That is to say, GANs can
utilize the numerical loss function to limit its adversarial process.


\subsection{Subjective Evaluation}
\label{ssec:subeva}
We conducted listening tests to assess the quality of the synthesized speech. we made four pairs of A/B preference
test: BLSTM vs. GAN, GAN vs. ASV, GAN vs. GAN-PC and BLSTM vs. GAN-PC. For listening test, 20 sentences
were randomly selected from the test data, and all listening pairs were presented in a shuffled order. There were 20
listeners participating in the test. In each test session, the listeners were asked to choose the better one considering
the perceived speech quality, or choose the ``neutral'' option if there was no difference.
\begin{figure}[htb]

\begin{minipage}[b]{1.0\linewidth}
  \centering
  \centerline{\includegraphics[width=8.5cm]{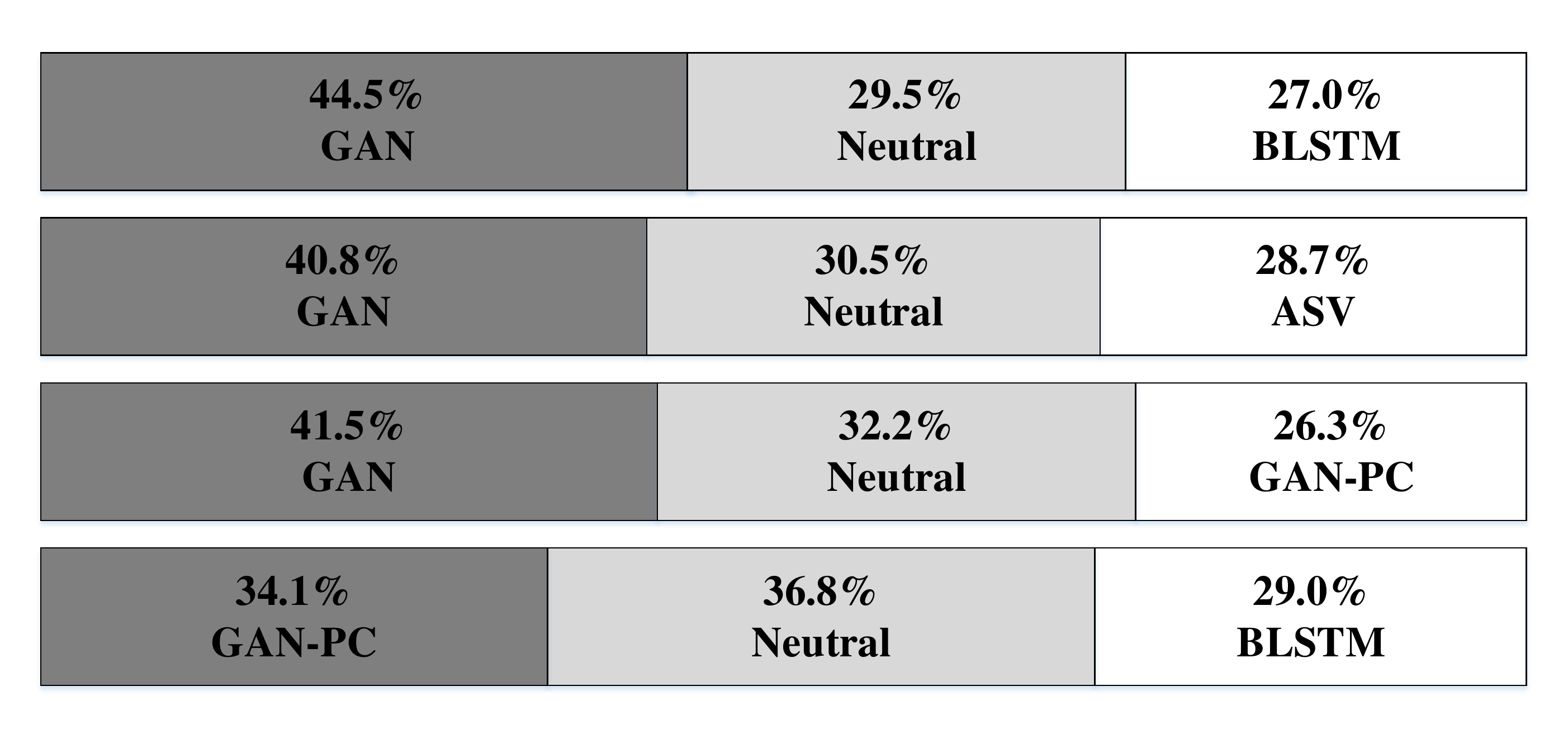}}
\end{minipage}
\caption{The preference score (\%) of A/B test.}
\label{fig:subjective}
\end{figure}

\begin{figure}[htb]

\begin{minipage}[b]{1.0\linewidth}
  \centering
  \centerline{\includegraphics[width=8.5cm]{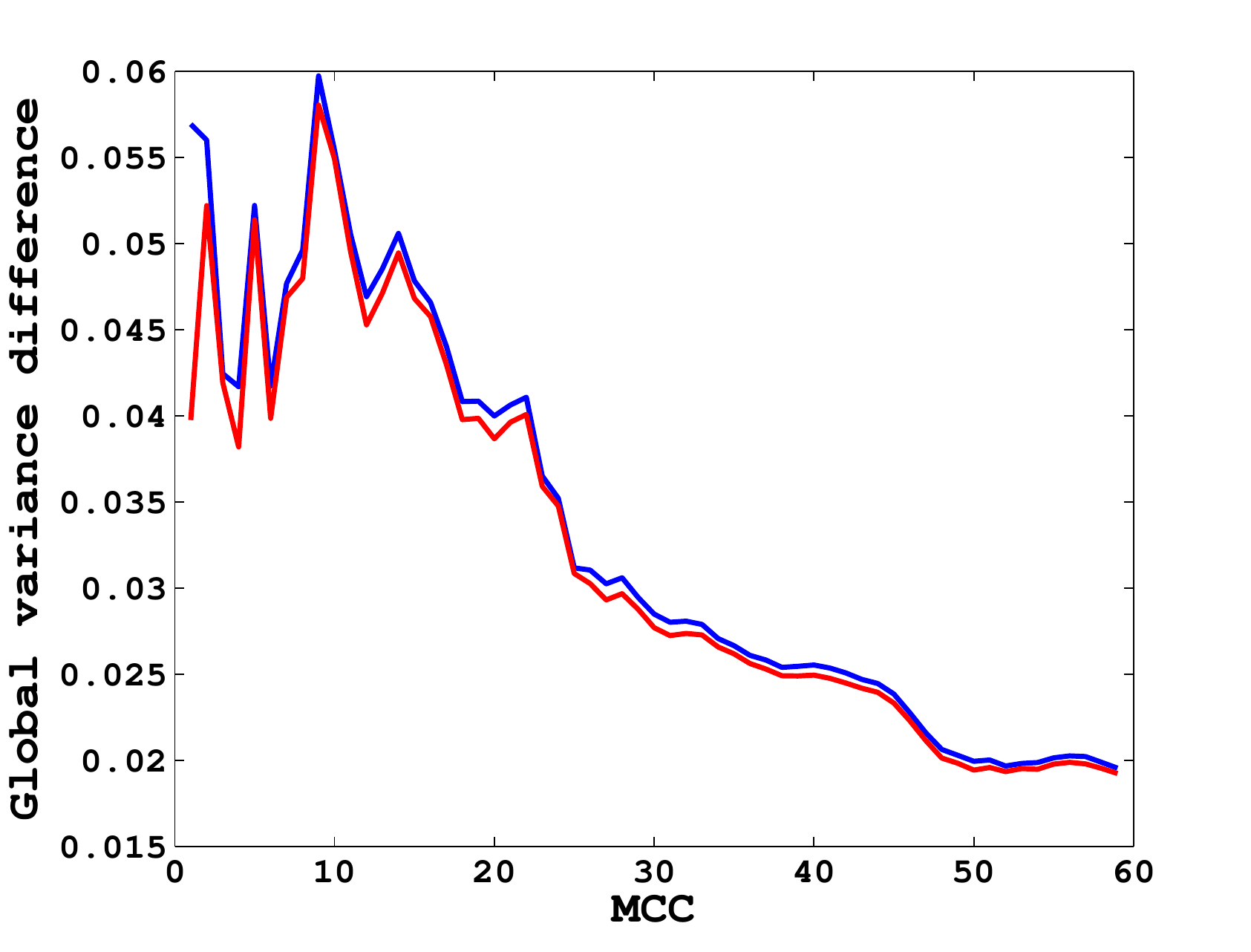}}
\end{minipage}
\caption{The difference of averaged GVs per mel-cepstral coefficients compared to natural speech.}
\label{fig:gv}
\end{figure}

Fig.~\ref{fig:subjective} shows the preference bars of the four pairs. The first bar of \textbf{GAN} vs. \textbf{BLSTM} indicates that the GAN-based
MTL framework can significantly improve the performance of the synthesized speech ($p<0.0001$). The listeners pointed out that the \textbf{GAN} system could produce speech with less buzzy sounds in most cases and more natural prosody in some samples.
The second bar of \textbf{GAN} vs. \textbf{ASV} shows that the proposed \textbf{GAN} approach is better than the \textbf{ASV} approach ($p < 0.0001$). As discussed in~\cite{saito2017training}, the ASV optimization aims to make the distribution of the synthesized speech close to the natural speech. But this method theoretically lacks the linguistic conditional guidance in distinguishing between the distributions of natural and synthesized speech. As a result, as compared with \textbf{ASV}, we find that the proposed \textbf{GAN} approach can capture both subtle and rapid changes, leading to better brightness of the synthesized speech.
Fig.~\ref{fig:gv} shows the distance of averaged global variance (GV) between natural speech and synthesized speech from different approaches. The smaller values mean that the GVs of synthesized speech are more similar to natural speech. The result indicates that the GVs of \textbf{GAN} are closer to the natural GVs than \textbf{AVS} especially in the first few coefficients.

The third bar of \textbf{GAN} vs. \textbf{GAN-PC} in Fig.~\ref{fig:subjective} shows that there is a decline in performance with phoneme classification in GANs ($p<0.001$).  In order to explain this  phenomenon, we compared the \textbf{BLSTM} system with \textbf{GAN-PC}, as shown in the last bar. In this preference test,
\textbf{GAN-PC} slightly outperforms \textbf{BLSTM}, but the difference is not significant between the two systems ($p=0.0253$).
As we know, the phoneme information is related to the contents of speech, which highly correlates to the intelligibility of the synthesized samples.
The BLSTM based acoustic model can already produce speech with high intelligibility. However, the purpose of treating the phoneme label as a guidance for the discriminator in \textbf{GAN-PC} is to improve the
intelligibility, not to make the distribution of synthesized examples closer to the natural samples. So simply letting the discriminator distinguish
whether $x$ is a natural sample in \textbf{GAN} can make the synthesized speech be more related to human perception, resulting in better subjective listening performance.

\section{CONCLUSION}
\label{conclusion}

This paper proposed to use GANs to improve the quality of synthesized speech. We use a multi-task learning architecture
with GANs, where the GANs can compensate the deficiency of traditional MSE loss function while the  MSE can also help to solve the instability
of GANs. Evaluation results show that the proposed method can compensate the weakness of numerical loss function and improve the
performance of SPSS. The proposed framework has a little increase in the computation cost, compared to traditional acoustic models during the generation process, as the extra computation only comes from noise generation and feature concatenation.

In our future work, we will focus on improving the performance of the generator in GAN and try to use GAN in end-to-end speech synthesis~\cite{wang2017tacotron,mehri2016samplernn,oord2016wavenet,arik2017deep,sotelo2017char2wav}. Since the the MSE loss is
still used to stabilize the adversarial process in our framework, we attempt to find a self-stabilizing architecture to directly estimate the distribution
of synthesized speech, such as using Wasserstein GAN~\cite{arjovsky2017wasserstein} and VAE-GAN~\cite{rosca2017variational}.


\bibliographystyle{IEEEbib}
\bibliography{main}

\end{document}